\documentclass[nofootinbib,prd,superscriptaddress,twocolumn,showpacs]{revtex4}

\usepackage{footnote}
\usepackage[multiple]{footmisc}
\makesavenoteenv{itemize}    
\makesavenoteenv{enumerate} 
\makesavenoteenv{description}

\usepackage{amsmath}
\usepackage{amsfonts}
\usepackage{amssymb}
\usepackage{graphicx}
\usepackage[titletoc]{appendix}
\usepackage{color}
\usepackage{hyperref}
\usepackage{cleveref}
\usepackage[rightcaption]{sidecap}
\usepackage{comment}
\usepackage{soul}
\usepackage{cancel}

\usepackage{dcolumn}

\usepackage{longtable}
\usepackage{floatrow}
\usepackage{array}
\usepackage{ctable}
\usepackage{multirow}
\usepackage{siunitx}
\usepackage{tabularx}
\usepackage{booktabs}
\usepackage{supertabular}

\graphicspath{{Graphics/}}

\def\be{\begin{equation}}
\def\ee{\end{equation}}
\def\bea{\begin{eqnarray}}
\def\eea{\end{eqnarray}}

\definecolor{vividviolet}{rgb}{0.62, 0.0, 1.0}
\definecolor{amaranth}{rgb}{0.9, 0.17, 0.31}
\definecolor{palatinateblue}{rgb}{0.15, 0.23, 0.89}
\definecolor{brightpink}{rgb}{1.0, 0.0, 0.5}
\definecolor{cornflowerblue}{rgb}{0.39, 0.58, 0.93}
\definecolor{deepcarminepink}{rgb}{0.94, 0.19, 0.22}
\definecolor{radicalred}{rgb}{1.0, 0.21, 0.37}

\hypersetup{ linktoc=all,
    colorlinks, linkcolor={palatinateblue},
    citecolor={brightpink}, urlcolor={amaranth}
}

\begin{document}

\title{Mapping the redshift drift at various redshifts through cosmography}

\author{Anna Chiara Alfano}
\email{a.alfano@ssmeridionale.it}
\affiliation{Scuola Superiore Meridionale, Largo S. Marcellino 10, 80138 Napoli, Italy.}
\affiliation{Istituto Nazionale di Fisica Nucleare (INFN), Via Cinthia 9, 80138 Napoli, Italy.}

\author{Orlando Luongo}
\email{orlando.luongo@unicam.it}
\affiliation{Universit\`a di Camerino, Divisione di Fisica, Via Madonna delle carceri 9, 62032 Camerino, Italy.}
\affiliation{Department of Nanoscale Science and Engineering, University at Albany SUNY, Albany, NY 12222, USA.}
\affiliation{INAF, Osservatorio Astronomico di Brera, Milano, Italy.}
\affiliation{Al-Farabi Kazakh National University, Al-Farabi av. 71, 050040 Almaty, Kazakhstan.}

\begin{abstract}
The redshift drift provides a kinematic test of the cosmic expansion history through the slow time variation of the redshift of comoving sources. Motivated by the expected Sandage-Loeb measurements from future facilities, we investigate the drift within a cosmographic framework, modeling the Hubble rate through both a second-order Taylor expansion and a $(2,1)$ Pad\'e approximant. We constrain the cosmographic parameters $(H_0,q_0,j_0)$ by combining Pantheon+ and SH0ES type Ia supernovae with gamma-ray bursts and then examine the impact of adding baryon acoustic oscillation measurements from the second DESI data release. The resulting constraints are used to construct a mock Sandage-Loeb catalog, after which the analyses are repeated including the simulated drift data. In this way, we assess the internal consistency of the reconstructed background rather than perform an independent forecast. Accordingly, we find that, for the SNeIa+GRB analysis, the Taylor reconstruction is compatible at the $1\sigma$ level with the $\omega_0\omega_1$CDM scenario, whereas the Pad\'e parameterization improves the agreement of $q_0$ with the $\Lambda$CDM paradigm. Once DESI BAO data are included, the agreement with the reference background models weakens to the $2\sigma$ level. The addition of the mock Sandage-Loeb sample mainly tightens the bounds on $q_0$ and $j_0$, with moderate shifts in the central values. We finally compare the reconstructed redshift drift with the corresponding behavior predicted by the $\Lambda$CDM and $\omega_0\omega_1$CDM scenarios.
\end{abstract}



\maketitle
\tableofcontents

\section{Introduction}

The redshift drift represents the phenomenon where the redshift between an emission source and an observer slowly evolves due to the expansion of the Universe, providing a direct measurement of cosmic acceleration \cite{1962ApJ...136..319S, 1962ApJ...136..334M, 1998ApJ...499L.111L}. 

Future surveys such as the European Extremely Large Telescope (E-ELT) with the high-resolution spectrograph COsmic Dynamics EXperiment (CODEX) or the Square Kilometer Array (SKA) \cite{2008MNRAS.386.1192L, 2015aska.confE..27K} will be able to detect the drift in redshift through the shift of the spectroscopic velocity $\Delta v$. E-ELT will map the matter epoch using the absorption lines of the Lyman-$\alpha$ forest of quasars probing the redshift range,  $2\leq z\leq 5$, known as the {\it redshift desert} \cite{2007PhRvD..75f2001C, 2012PhRvD..86l3001M} while SKA will investigate the acceleration epoch considering observations of the emission in galaxies through the hydrogen emission lines in the range $0\leq z \leq 1$. These measurements will provide a direct test of the cosmic expansion history and possibly investigate the nature of dark energy \cite{2007MNRAS.382.1623B, 2010PhRvD..81d3522Q, 2012PhRvD..85h7301V, 2014JCAP...07..006G, 2014JCAP...12..018G, 2011PhRvD..84j4003M, 2015EPJC...75..356G, 2021PhRvD.103h1302H, 2022LRR....25....6M}. 

Recent DESI results \cite{2025JCAP...07..028A, 2025PhRvD.112h3515A} have revived the possibility that dark energy deviates from a cosmological constant \cite{1992ARA&A..30..499C, 2001LRR.....4....1C, 2003RvMP...75..559P}. In this context, probes directly tracing the expansion history provide a natural avenue to test this scenario. Indeed, since its first data release in 2024, many works have been proposed that dismiss or sustain DESI's claim of a dynamical dark energy, see e.g. Refs. \cite{2024A&A...690A..40L, 2024JCAP...12..055A, 2025PhRvD.111b3512C, 2025PhRvD.112f3523S, 2025arXiv250322529N, 2025PhRvD.112b3508G, 2025PhRvD.111l3511S, 2025MNRAS.542L..24O, 2025PhRvD.112j3519R, 2026JHEAp..4900428O, 2025arXiv251116130N}.

Here, we adopt a model-independent approach based on {\it cosmography} \cite{1972gcpa.book.....W, 1976Natur.260..591H, 2004CQGra..21.2603V, 2005GReGr..37.1541V, 2010dmap.conf..287V, 2011MPLA...26.1459L, 2016IJGMM..1330002D} that relies only on the validity of the cosmological principle. Cosmography provides information on kinematic quantities through the derivatives of the scale factor $a(t)$ by expanding it in a Taylor series around present time, up to a given order, yielding the so-called {\it cosmographic parameters}. However, even though this approach is adopted in many contexts, see e.g. Refs. \cite{2012arXiv1211.0626L, 2015arXiv151207076L, 2018JCAP...05..008C, 2019IJMPD..2830016C, 2021PhRvD.104h3519T, 2022JCAP...03..057H, 2022MNRAS.509.5399C, 2023PhRvD.108b3523D, 2023PDU....4201298A, 2025JCAP...01..088J, 2025arXiv250701095M, 2025arXiv251219568V, 2025JHEAp..4800405D}, its major drawback lies in the truncation of the Taylor expansion of $a(t)$ leading to convergence issues at $z\geq 1$. To circumvent this problem, various methods have been proposed throughout the years from auxiliary variables which reparameterize the redshift \cite{2007CQGra..24.5985C, 2012PhRvD..86l3516A} to rational polynomials such as Pad\'e \cite{2014PhRvD..90d3531A, 2014PhRvD..89j3506G, 2020MNRAS.494.2576C} or Chebyshev  \cite{2018MNRAS.476.3924C}.

In this work, we consider the first three cosmographic parameters, $H(t)$, $q(t)$, and $j(t)$. The sign of $q(t)$ determines acceleration, while $j(t)>0$ signals the transition to dark energy domination. In the $\Lambda$CDM concordance model, $j_0=1$, so deviations from this value may point to non-standard dark energy.

Our strategy is therefore to combine multiple cosmological probes to map redshift regions from local to earlier epochs together with the so-called {\it Sandage-Loeb} test\footnote{It offers a direct test of the expansion history of the Universe by considering measurements of the evolution of cosmic redshift through the use of quasar's spectra \cite{2007PhRvD..75f2001C}.} which adopts as an observable the velocity shift $\Delta v$ of a source in a time interval $\Delta t$ to constrain $H_0$, $q_0$ and $j_0$ through 
\begin{equation}\label{vshift}
    \Delta v\equiv \frac{c\Delta z}{1+z}=cH_0\Delta t_0\left[1-\frac{H(z)}{H_0(1+z)}\right],
\end{equation}
where $c$ is the speed of light, $\Delta t_0$ is the time interval of observations, $H_0$ is the Hubble constant and $H(z)$ the Hubble rate written in terms of a i) Taylor expansion and of a ii) Pad\'e approximant of order $(2, 1)$.

The datasets adopted in this work are type Ia supernovae (SNeIa) from the Pantheon+ sample \cite{2022ApJ...938..113S} calibrated with Cepheids from the SH0ES program \cite{2022ApJ...934L...7R}, gamma-ray burst (GRB) data fulfilling the $E_p-E_{iso}$ or {\it Amati} correlation function \cite{2021JCAP...09..042K}, the baryonic acoustic oscillations (BAO) from the second data release (DR2) by the DESI Collaboration \cite{2025PhRvD.112h3515A} and the Sandage-Loeb (SL) mock data catalog of the velocity shifts. 

Remarkably, in our computations we first standardize GRB to use them as distance indicators. Specifically, we adopt the well-established method based on Bézier parametric curves where $H(z)$ is approximated through B\'ezier polynomials \cite{2019MNRAS.486L..46A, 2021MNRAS.501.3515M, 2021MNRAS.503.4581L, 2023MNRAS.523.4938M, 2023MNRAS.518.2247L, 2024JCAP...12..055A, 2025JHEAp..4600348A, 2025A&A...701A.220L} adopting observational Hubble data (OHD) to make the luminosity distance appearing in the radiated isotropic energy $E_{iso}$ model-independent.

The adoption of GRBs together with SL data will strengthen the constraints on the parameters since both map the interval $2\leq z \leq 5$ with GRBs pushing over $z\geq 5$. The constraints on the parameters are derived by modifying the Cosmic Linear Anisotropy Solving System (\texttt{CLASS}) code \cite{2011arXiv1104.2932L, 2011JCAP...07..034B, 2011arXiv1104.2934L} and performing a Monte Carlo Markov chain (MCMC) analysis adopting the \texttt{MontePython} sampler \cite{2019PDU....24..260B, 2013JCAP...02..001A}.

Specifically, we follow the strategy outlined in Refs. \cite{2014JCAP...07..006G, 2014JCAP...12..018G}, precisely,

\begin{itemize}
    \item [-] we do preliminary MCMC runs to constrain the Hubble, deceleration and jerk parameters with the adoption of SNeIa+GRB (\emph{Analysis 1}) and with SNeIa+GRB+DESI-BAO (\emph{Analysis 2}) to assess the impact of the BAO data on the parameters;
    \item [-] we substitute for each analysis the constrained parameters inside Eq. \eqref{vshift} to generate the mock SL data catalogs;
    \item [-] we perform the MCMC analyses again for every combination of datasets with the inclusion of the SL probe to assess how it affects the fits carrying out a consistency test. In this case, we label them \emph{Analysis 1SL} and \emph{Analysis 2SL}.
\end{itemize}

Considering the preliminary MCMC our inferred parameters are compared with $H_0$ from the Planck Collaboration \cite{2020A&A...641A...6P} ($H^P_0= 67.36 \pm 0.54\ \text{km/s/Mpc}$) and from \citet{2022ApJ...934L...7R} ($H^R_0= 73.04\pm 1.04\ \text{km/s/Mpc}$). Then, we compare the inferred deceleration parameters $q_0$ and the jerk parameters $j_0$ with those derived by considering as background the $\Lambda$CDM and $\omega_0\omega_1$CDM scenarios.

Our preliminary results suggest that, when we do not consider DESI data, i.e. \emph{Analysis 1} the Hubble constant $H_0$ inferred from our analysis agrees at $1$-$\sigma$ with $H^R_0= (73.04\pm 1.04)\ \text{km/s/Mpc}$ \citet{2022ApJ...934L...7R} while it is not compatible with $H^P_0= (67.36 \pm 0.54)\ \text{km/s/Mpc}$ \cite{2020A&A...641A...6P}. This is found for both the adoption of the Hubble rate written in terms of Taylor or Pad\'e series. As for the deceleration and jerk parameters, when Taylor is adopted $q_0$ and $j_0$ agree at $1$-$\sigma$ only with the values from the $\omega_0\omega_1$CDM model. Switching to Pad\'e, both $q_0$ and $j_0$ now also agree at $1$-$\sigma$ with the concordance paradigm.

When DESI data are considered, i.e. \emph{Analysis 2} we find that now our $H_0$ does not agree with either the value from the Planck Collaboration \cite{2020A&A...641A...6P} or from \citet{2022ApJ...934L...7R} in both scenarios of adopting $H(z)$ written in terms of Taylor or Pad\'e. This is also the case of the jerk parameter $j_0$ which finds no agreement with the prediction from both $\Lambda$CDM and $\omega_0\omega_1$CDM models. On the other hand, the constraints on the deceleration parameter $q_0$ are the same as in \emph{Analysis 1} when using Taylor while the agreement degrades to the $2$-$\sigma$ with both background cosmologies when Pad\'e is used.

Importantly, these preliminary results on $j_0$ agree with other works in the literature. For example, \citet{2024A&A...690A..40L} adopt the first data release (DR1) of the DESI Collaboration \cite{2025JCAP...07..028A} in combination with OHD first, SNeIa and a combination of DESI-DR1+OHD+SNeIa assessing a tension between their inferred values of $j_0$ and the jerk parameter from the $\Lambda$CDM and $\omega_0\omega_1$CDM models. Recently, \citet{2025PhRvD.112j3519R} found that when employing DESI-DR2 together with various supernova samples $j_0$ deviates from $1$ at $3.4\sigma$ (DESI-DR2+Union3), $4.1\sigma$ (DESI-DR2+Pantheon+) and $5.4\sigma$ (DESI-DR2+DESY5).

Afterwards, we conclude our MCMC computations by including a constructed SL mock data catalog labeling these additional analyses as \emph{Analysis 1SL} and \emph{Analysis 2SL}, respectively. Firstly, we observe a reduction of the errors on $q_0$ and $j_0$ in both \emph{Analysis 1SL}-\emph{Analysis 2SL} and only minor variations in the results.

Specifically, for \emph{Analysis 1SL} where SNeIa+GRB+SL are considered, when using Taylor we observe a drop in compatibility when the $\omega_0\omega_1$CDM is considered changing from $1$-$\sigma$ to $2$-$\sigma$. The same can be seen for the jerk parameter $j_0$ which is now not compatible anymore with $1$ from the $\Lambda$CDM scenario. This changes when the Pad\'e approximant is adopted, since in this case the jerk parameter is also compatible with $j_0=1$ at the $1\sigma$ level. When DESI data are taken into account, i.e. \emph{Analysis 2SL} when Pad\'e is used there is no more compatibility between our $q_0$ and the one from the concordance paradigm while it remains at $2$-$\sigma$ for the $\omega_0\omega_1$CDM model.

In view of these results, we want to stress that the change in compatibility when adding the mock SL data is mainly observed in the deceleration parameter $q_0$ when DESI is used. This is not surprising since the SL signal is a probe adopted to investigate the acceleration or deceleration of the Universe which is encoded in the deceleration parameter. However, when we drop the BAO sample we observe a shift in compatibility also on the jerk parameter $j_0$. The reduction of the errors for most parameters when adopting the mock SL data together with tighter confidence regions as shown in Fig. \ref{fig:contSL} is in part due to the adoption of the same approach used to generate the preliminary constraints on the cosmographic parameters, i.e. the adoption of the Hubble rate $H(z)$ written using the Taylor or Pad\'e series.

Finally, we also plot the behavior of the drift using the parameters found in \emph{Analysis 1SL-2SL} comparing them with the behavior of the drift considering the concordance paradigm and the $\omega_0\omega_1$CDM model.

The present work is organized as follows. In Sect. \ref{cosmography} we outline the cosmographic approach and introduce the Hubble rate written using the Taylor and Pad\'e approaches. Afterwards, in Sect. \ref{rshift} we present the redshift drift starting from its derivation to its form when adopting the cosmographic approach. Sect. \ref{cosprob} deals with illustrating the cosmic probes adopted in this work focusing also on possible biases in BAO data while in Sect. \ref{numout} we show the preliminary numerical outcomes of the cosmographic parameters that will be used to construct the mock SL data catalog. Sect. \ref{mocksl} shows the final results from our MCMC computations with the addition of the SL data together with showing the behavior of the $z$-drift from the inferred cosmographic parameters comparing it with the $\Lambda$CDM and $\omega_0\omega_1$CDM scenarios. Finally, in Sect. \ref{conc} we present the conclusions drawn from our work.

\section{The cosmographic approach}\label{cosmography}

The only assumption made in cosmography is the validity of the cosmological principle. This provides a powerful method to investigate the expansion history of the Universe without assuming {\it a priori} a cosmological model. In particular, the cosmographic parameters are found by expanding the scale factor $a(t)$ around the present time $t_0$
\begin{equation}
    a(t)=1+\sum^{\infty}_{n=1}\frac{1}{n!}\frac{d^na}{dt^n}\Bigg|_{t=t_0}(t-t_0)^n,
\end{equation}
where up to the third order we have the Hubble, deceleration and jerk parameters \cite{2005GReGr..37.1541V, 2004CQGra..21.2603V, 2010dmap.conf..287V, 2011MPLA...26.1459L, 2016IJGMM..1330002D}
\begin{subequations}

\begin{align}\label{hubble}
    H(t)=\frac{1}{a}\frac{da}{dt},\\
    q(t)=-\frac{1}{aH^2}\frac{d^2a}{dt^2},\\
    j(t)=\frac{1}{aH^3}\frac{d^3a}{dt^3},
\end{align}
\end{subequations}
where the Hubble rate $H(t)$ needs to be positive for the Universe to expand while the deceleration and the jerk parameters can be separated into three cases each. 

Focusing on $q(t)$ first at present times it can be i) $q_0 >0$ indicating a Universe dominated by a pressureless barotropic fluid ii) $-1<q_0<0$ corresponding to a speeding up Universe and iii) $q_0=-1$ implying that the Universe is dominated by a de-Sitter phase \cite{2014PhRvD..90d3531A}.

On the other hand, the possible values that $j(t)$ can assume are directly linked through the variation of $q(t)$
\begin{equation}
    j=(1+z)\frac{dq}{dz}+2q^2+q.
\end{equation}

At present times $j$ can be i) $j_0<0$ hinting at a dark energy fluid influencing the dynamics at early times ii) $j_0=0$ assuming that $q(t)$ approaches a precise value as the redshift tends to infinity iii) $j_0>0$ forecasting a precise redshift labeled as transition redshift at which the deceleration parameter changes sign and the dark energy-dominated phase begins \cite{2014PhRvD..90d3531A}. Additionally, it was recently pointed out that $j_0>1$ may suggest a freezing quintessence scenario where the field freezes as it slows down mimicking a cosmological constant at later epochs \cite{2025PhRvD.112j3519R}.

It is useful to write the Hubble rate in terms of the aforementioned cosmographic parameters by first expanding $H(z)$ in a Taylor series \cite{2025arXiv250916196F,Bamba:2012cp}
\begin{equation}
    H(z)=H(z_0)+H^\prime(z_0)(z-z_0)+\frac{1}{2}H^{\prime\prime}(z_0)(z-z_0)^2+\mathcal{O}(z-z_0)^3,
\end{equation}
where the $^\prime$ denotes the derivative with respect to the redshift $z$. Then, starting from Eq. \eqref{hubble} and taking the derivatives of the Hubble rate at present times, i.e. $z_0\equiv 0$ we end up with
\begin{equation}\label{hcosmograph}
    H_T(z)\simeq H_0\left[1+(1+q_0)z+\frac{1}{2}(j_0-q_0^2)z^2\right].
\end{equation}
Other than using a Taylor expansion on the Hubble rate we will also investigate the form of $H(z)$ in terms of the so-called Pad\'e polynomials aiming at alleviating possible convergence problems arising at high redshifts \cite{2014PhRvD..90d3531A, 2018JCAP...05..008C, 2019IJMPD..2830016C}
\begin{equation}
H_{21}(z)=2H_0(1+z)^2\,\frac{A_0+A_1 z}{B_0+B_1 z+B_2 z^2},
\label{hpade}
\end{equation}
with
    \begin{subequations}
\begin{align}
A_0 &= 3(1-q_0),\\
A_1 &= j_0+1-q_0-3q_0^2,\\
B_0 &= 18(q_0-1)^2,\\
B_1 &= 6(q_0-1)\bigl(q_0(8+3q_0)-5-2j_0\bigr),\\
B_2 &= 14+j_0(7+2j_0)-q_0\big(j_0(10+9q_0)+\nonumber\\
&q_0(-40+q_0(17+9q_0(2+q_0))\big).
\end{align}
\end{subequations}
\section{The redshift drift}\label{rshift}

The basic concept behind the redshift drift is that in an expanding Friedmann-Lema\^itre-Robertson-Walker (FLRW) Universe the redshift measured between an emitter and an observer evolves slowly with cosmic time \cite{1962ApJ...136..319S, 1962ApJ...136..334M, 1998ApJ...499L.111L}. To derive it we follow the derivation outlined in Ref. \cite{2020JCAP...04..043L}. We consider the scale factor $a(t_0)$ at the current epoch and at the time of the emission event $a(t_e)$ so that 
\begin{equation}\label{z}
    1+z=\frac{a(t_0)}{a(t_e)}.
\end{equation}
Taking the derivative with respect to the time of Eq. \eqref{z} we end up with
\begin{equation}
    \Dot{z}= \frac{\Dot{a}(t_0)a(t_e)-a(t_0)(da(t_e)/dt_0)}{a^2(t_e)}.
\end{equation}
Adopting the chain rule and considering that $\Dot{a}(t_0)/a(t_0)\equiv H_0$ at present times leads us to \cite{1962ApJ...136..334M}
\begin{equation}\label{zdrift}
    \Dot{z} = (1+z)H_0-H(z),
\end{equation}
with $H_0$ being the Hubble constant at current epoch and $H(z)$ the Hubble rate.

The use of the redshift drift provides a powerful discriminator to determine  if the Universe is accelerating or decelerating. Precisely, a positive $\Dot{z}$ hints at a violation of the strong energy condition within a FLRW framework which in the standard $\Lambda$CDM scenario indicates a fluid with an equation of state $\omega\equiv p/\rho=-1$ speeding up the Universe at present times \cite{2021PhRvD.103h1302H}.

This can be seen more clearly when the redshift drift is written adopting the cosmographic approach \cite{2020JCAP...04..043L}. Specifically, we take the Hubble rate written in terms of the cosmographic parameters in Eq. \eqref{hcosmograph} and substitute it inside Eq. \eqref{zdrift}, yielding
\begin{equation}\label{zdriftcosmo}
    \Dot{z}_T \simeq -H_0\left[q_0z+\frac{1}{2!}(j_0-q_0^2)z^2\right].
\end{equation}
On the other hand, adopting Pad\'e polynomials $\Dot{z}$ will take the following form 
\begin{equation}
\dot z_{21}=H_0(1+z)\,\frac{\mathcal N(z)}{\mathcal D(z)}\,,
\label{zdriftpade}
\end{equation}
with
\begin{widetext}
\begin{subequations}
    \begin{align}
\mathcal N(z) &= 6\bigl[1+3(q_0-1)\bigr]
+6(q_0-1)\bigl[q_0(8+3q_0)-2(3+j_0)-2(j_0+1-3q_0^2)\bigr]z
\nonumber\\
&\quad
+\bigl[14+j_0(7+2j_0)-q_0\bigl(50+q_0(9q_0(2+q_0)-8)-j_0-1\bigr)\bigr]z^2,
\\[1ex]
\mathcal D(z) &= 6(q_0-1)\bigl[3(q_0-1)+(q_0(8+3q_0)-5-2j_0)z\bigr]
\nonumber\\
&\quad
+\bigl[14+j_0(7+2j_0)-q_0\bigl(10+9q_0+40-q_0(17+9q_0(2+q_0))\bigr)\bigr]z^2.
\end{align}
\end{subequations}
\end{widetext}
Since in the concordance paradigm $q_0=-0.527$ and $j_0=1$ it is straightforward to check that $\Dot{z}>0$ in both Eqs. \eqref{zdriftcosmo}-\eqref{zdriftpade}.

To investigate the redshift drift via the cosmographic parameters, we first

\begin{itemize}
    \item [i)] consider the Hubble rate $H(z)$ written adopting various parameterizations as seen in Eqs. \eqref{hcosmograph}-\eqref{hpade}.
    \item [ii)] substitute the parameterizations of $H(z)$ introduced in point i) into the luminosity distance, in order to constrain the cosmographic parameters.
    \begin{equation}\label{dlum}
        d_L(z) = \frac{c(1+z)}{H_0\sqrt{\Omega_k}}\sinh\left[\int^z_0\frac{H_0\sqrt{\Omega_k}}{H^\star(z^\prime)}\ dz^\prime\right],
    \end{equation}

    where $c$ is the speed of light, $H^\star(z)$ varies according to whether we are considering the Taylor or Pad\'e approximations while $\Omega_k$ is the curvature parameter which we consider $\Omega_k=0$, working in a flat scenario. 
    \item [iii)] we use the derived cosmographic parameters to construct a mock data catalog of the observable of the $z$-drift, the velocity shift $\Delta v$ considering 
\begin{equation}\label{vshift1}
    \Delta v= cH_0\Delta t_0\left[1-\frac{H^\star(z)}{H_0(1+z)}\right],
\end{equation}
with $H^\star(z)$ being Eqs. \eqref{hcosmograph}-\eqref{hpade} where we substitute inside them the cosmographic parameters inferred from our computations and $\Delta t_0$ the time of the observation. Following the approach proposed in Refs. \cite{2014JCAP...12..018G, 2014JCAP...07..006G} we repeat the computations outlined in ii) with the addition of the mock SL catalog to establish how they affect the analyses.
\end{itemize}

Finally, with the constrained cosmographic parameters derived in iii) we substitute them inside Eqs. \eqref{zdriftcosmo}-\eqref{zdriftpade} and plot the results for the various parameterizations employed in this work considering as baseline cosmological models the concordance paradigm and the $\omega_0\omega_1$CDM scenario.

\section{Cosmic probes}\label{cosprob}

The cosmic probes used in our analyses encompass both low-, intermediate- and high-$z$ regimes.

\begin{itemize}
    \item [-] {\bf Pantheon+ and SH0ES.} We consider the largest SNeIa catalog consisting of 1701 data points within the redshift range $z\in[10^{-3}, 2.3]$ together with the SNeIa host galaxy distances calibrated through the Cepheids encapsulated in the SH0ES sample providing constraints on $H_0$ \cite{2022ApJ...938..113S, 2022ApJ...938..110B}.

    With the adoption also of the SH0ES sample the distance modulus residuals $\Delta\mu_i$ take the following form
    \begin{equation}
        \Delta\mu_i=\begin{cases}
        \mu_i-\mu^C_i, \quad &\text{with $i\in\text{Cepheid hosts}$},\\
        \mu_i-\mu(z_i)\quad &\text{otherwise},
        \end{cases}
    \end{equation}
    where $\mu_i$ is the distance modulus measured by SNeIa, $\mu^C_i$ represents the host galaxy distance modulus values calibrated using Cepheids while $\mu(z_i)$ denotes the theoretical distance modulus defined as 
    \begin{equation}\label{mudis}
        \mu(z_i)=25+5\log\left[\frac{d_L(z_i)}{\text{Mpc}}\right],
    \end{equation}
   with the luminosity distance provided by Eq. \eqref{dlum}. The log-likelihood for this sample is given by
   \begin{equation}
       \ln\mathcal{L}_S=-\frac{1}{2}\Delta\mu^T\mathcal{C}_S^{-1}\Delta\mu,
   \end{equation}
   where $\Delta\mu=\mu_i-\mu(z_i)$ with $\mu_i$ representing the data points in the sample while $\mu(z_i)$ the theoretical distance moduli. The covariance matrix $\mathcal{C}_S=\mathcal{C}_{\text{stat}}+\mathcal{C}_{\text{sys}}$ takes into account both the statistical and systematic errors\footnote{The complete covariance matrix can be found in the \texttt{GitHub} repository \url{https://github.com/PantheonPlusSH0ES/DataRelease}.}. 

   \item [-] {\bf DESI DR2.} We adopt the second data release of the DESI Collaboration \cite{2025PhRvD.112h3515A} consisting of 13 data points in the redshift interval $z\in [0.295,\ 2.33]$ as presented in Tab. IV of Ref. \cite{2025PhRvD.112h3515A} involving measurements of the Hubble rate distance $d_H/r_d$, the transverse comoving distance $d_M/r_d$ and the angle-averaged distance $d_V/r_d$
   \begin{subequations}
       \begin{align}
          & \frac{d_H(z)}{r_d}=\frac{c}{r_dH^\star(z)},\\
           &\frac{d_M(z)}{r_d}=\frac{d_L(z)}{r_d(1+z)},\\
           &\frac{d_V(z)}{r_d}=\frac{d^{2/3}_M(z)[zd_H(z)]^{1/3}}{r_d},
       \end{align}
   \end{subequations}
   where $H^\star(z)$ are Eqs. \eqref{hcosmograph}-\eqref{hpade}. Further, we fix the sound horizon at the drag epoch to the value inferred by the Planck Collaboration \cite{2020A&A...641A...6P} $r_d=(147.09\pm 0.26)\ \text{Mpc}$ to break the degeneracy between $r_d$ and $H_0$.

   The log-likelihood for this dataset is
   \begin{equation}
       \ln\mathcal{L}_D=-\frac{1}{2}\Delta D^T\mathcal{C}^{-1}_D\Delta D,
   \end{equation}
   where $\Delta D=D_i-D(z_i)$ with $D_i=\{d_H/r_d,\ d_M/r_d,\ d_V/r_d\}$ the data points from the second DESI data release while $D(z_i)$ are the theoretical distances varying according to the $H^\star(z)$ chosen through Eq. \eqref{dlum}. Moreover, $\mathcal{C}_D$ is the covariance matrix of the sample\footnote{The complete covariance matrix can be found in the \texttt{GitHub} repository \url{https://github.com/CobayaSampler/bao_data/blob/master/desi_bao_dr2/desi_gaussian_bao_ALL_GCcomb_cov.txt}}.

    \item [-] {\bf GRBs.} We employ these powerful astrophysical sources that can be found at redshifts as high as $z\simeq 9$  \cite{2009Natur.461.1258S, 2011ApJ...736....7C} through various correlation functions connecting different observables \cite{2002A&A...390...81A, 2006MNRAS.372..233A, 2004ApJ...609..935Y, 2004ApJ...616..331G, 2008MNRAS.391L..79D, 2015A&A...582A.115I}. 
    
    In this work, we adopt 118 GRBs following the $E_p-E_{iso}$ or {\it Amati} correlation spanning a redshift interval of $z\in[0.34, 8.2]$ \cite{2021JCAP...09..042K}. The $E_p-E_{iso}$ correlation combines the peak energy in the rest-frame $E_p$ and the burst's isotropic radiated energy $E_{iso}$ \cite{2002A&A...390...81A, 2006MNRAS.372..233A}
    \begin{equation}\label{amati}
        \log\left(\frac{E_p}{\text{keV}}\right)=b+a\left[\log\left(\frac{E_{iso}}{\text{erg}}\right)-52\right],
    \end{equation}
    where $a$ is the slope and $b$ the intercept.

    Although powerful for investigating intermediate/high-$z$ regions, they suffer from the so-called {\it circularity} problem jeopardizing their use as standard candles since the shortage of GRBs at low-$z$ prevents them from being anchored to primary distance indicators \cite{2021Galax...9...77L}. Specifically, through their observables they carry the luminosity distance which depends on an {\it a priori} cosmological model. In the specific case of the $E_p-E_{iso}$ correlation the dependence on $d_L$ appears in
    \begin{equation}
        E_{iso}(z)=4\pi d^2_L(z)S_b(1+z)^{-1},
    \end{equation}
    where $S_b$ is the bolometric fluence in units of erg/$\text{cm}^2$/s.
    
    To circumvent this issue many approaches have been proposed throughout the years, see e.g. Refs. \cite{2011MNRAS.415.3580D, 2015GReGr..47..141L, 2023MNRAS.521.4406L, 2021MNRAS.501.3515M, 2025Ap&SS.370...10Z} where the reconstruction of $H(z)$ through B\'ezier polynomials seems to be one of the most promising approaches \cite{2019MNRAS.486L..46A, 2021MNRAS.501.3515M, 2021MNRAS.503.4581L, 2023MNRAS.523.4938M, 2023MNRAS.518.2247L, 2024JCAP...12..055A, 2025JHEAp..4600348A, 2025A&A...701A.220L}.

    The complete log-likelihood for the GRB sample is $\ln\mathcal{L}_A=\ln\mathcal{L}^{\text{cal}}_A+\ln\mathcal{L}^{\text{cos}}_A$ where
    \begin{subequations}
        \begin{align}
            &\ln\mathcal{L}^{\text{cal}}_A\propto-\frac{1}{2}\sum^{N^{\text{cal}}_A}_{i=1}\left[\frac{\log E_{p,i}-\log E_p(z_i)}{\sigma_i}\right]^2,\\ \label{lncosmo}
            &\ln\mathcal{L}^{\text{cos}}_A\propto-\frac{1}{2}\sum^{N^{\text{cos}}_A}_{i=1}\left[\frac{\mu_i-\mu(z_i)}{\sigma_{\mu,i}}\right]^2,
        \end{align}
    \end{subequations}
    where $\log E_p(z_i)$ is defined in Eq. \eqref{amati} with $\log E_{iso}(z_i)$ derived by using the calibration procedure outlined in Appendix \ref{calibration}  and $\sigma^2_i=\sigma^2_{\log E_{p,i}}+a^2\sigma^2_{\log E_{iso,i}}+\sigma^2_{\text{ext}}$ with $\sigma_{\text{ext}}$ being an extra-scatter term \cite{2005physics..11182D}. The distance moduli depicted in Eq. \eqref{lncosmo} is
    \begin{equation}
        \mu_i=\frac{5}{2}\left[\frac{\log E_{p,i}-b}{a}-\log\left(\frac{4\pi S_{b,i}}{1+z_i}\right)\right]
    \end{equation}
    and $\sigma^2_{\mu,i}=25/4\left((\sigma^2_{\log E_{p,i}}+\sigma_{\text{ext}}^2)/a^2+\sigma^2_{\log S_{b,i}}\right)$ where also in this case $\sigma_{\text{ext}}$ is the same extra-scatter term appearing in $\sigma^2_i$. The theoretical distance moduli takes the form as in Eq. \eqref{mudis}.

    \item [-] {\bf Mock Sandage-Loeb.} We generate a total of $30$ measurements of the velocity shift equally split in the redshift interval $z\in [2, 5]$ and distributed in the redshift bins $z_{SL}=[2.0, 2.8, 3.5, 4.2, 5.0]$ \cite{2012PhRvD..86l3001M} for a total of $6$ data in each bin. The data points are derived considering Eq. \eqref{vshift1} with the values of $H_0$, $q_0$ and $j_0$  taken from {\it Analyses 1-2} displayed in Tab. \ref{tab:bfcosmo} and with a time for observations of $\Delta t_0=30\ \text{yrs}$. The associated error with each measurement in units of cm/s is \cite{2008MNRAS.386.1192L}
    \begin{equation}\label{strumerr}
        \sigma_{\Delta v}=1.35\left(\frac{2370}{S/N}\right)\sqrt{\frac{30}{N_{SL}}}\left(\frac{5}{1+z_{SL}}\right)^{1.7-x},
    \end{equation}
    where the signal-to-noise ratio $S/N$ is taken to be equal to $3000$ while $N_{SL}$ and $z_{SL}$ represent the number of quasars and the redshift at which they are found. Additionally, $x$ varies according to the redshift: it is $x=0$ when $z\leq 4$ and $x=0.8$ when $z> 4$. 

    However, since in our computations for the mock data catalog we need to take into account the cosmographic parameters with attached errors from the preliminary MCMC the total error becomes
    \begin{equation}
     \sigma^{\text{tot}}_{\Delta v}=\sqrt{\left(\frac{\partial \Delta v}{\partial A}\right)^2\sigma^2_A+\sigma^2_{\Delta v}},   
    \end{equation}
    where $A=\{H_0,\ q_0,\ j_0\}$ are the cosmographic parameters with attached errors $\sigma_A$ while $\sigma_{\Delta v}$ is simply Eq. \eqref{strumerr}.
    
    The log-likelihood in this case will be 
    \begin{equation}
        \ln\mathcal{L}_{SL}\propto-\frac{1}{2}\sum^{N_{SL}}_{i=1}\left[\frac{\Delta v_i-\Delta v(z_i)}{\sigma^{\text{tot}}_{\Delta v_i}}\right]^2,
    \end{equation}
    where $\Delta v_i$ are the generated data points with attached $\sigma_{\Delta v_i}$ errors while $\Delta v(z_i)$ are the theoretical velocity shift in Eq. \eqref{vshift1}. It is worth stressing that even though treating the data as correlated, i.e. defining a covariance matrix would be more realistic we treat them as uncorrelated due to lack of information \cite{2018EPJC...78...11L}. Specifically, to build a covariance matrix for this data we should take into consideration, e.g. whether the surveys consider overlapping redshift bins or not \cite{2018EPJC...78...11L}.
\end{itemize}

\subsection{Problems with baryonic acoustic oscillation data}

In the early Universe, in the plasma formed by baryons and photons, baryonic matter was attracted towards the structures formed by dark matter. At the same time, when matter fell inside these structures, an outward pressure counterbalanced the effect of gravity and brought baryons apart from dark matter. As the Universe cooled down due to the expansion, gravity started to pull again together baryons and dark matter creating sound waves that spread outward in bubble dragging matter in them imprinting in the large-scale structures through the scale of the sound horizon. This process formed the BAO appearing in the form of ripples in the clustering power spectrum and acoustic peak in the 2-point correlation function at a separation of $150\ \text{Mpc}$ \cite{2005ApJ...633..560E}.

This acoustic peak is a powerful standard ruler, adopted to map the expansion history of the Universe. However, as pointed out in \citet{2019PhRvD..99l3515A}, extracting distances through the correlation function is not a model-independent process. Specifically, due to the sensitivity of the distribution of matter at late times to non-linear dynamics of clustering, the peak of the BAO is shifted and lowered \cite{2007ApJ...664..660E, 2018PhRvD..98b3527A}. This requires fitting the correlation function, depending on a background cosmological model \cite{2009MNRAS.400.1643S, 2012MNRAS.425..415S} with fiducial values of the cosmological parameter to fix the BAO peak and deriving the distances leading to possible underestimation of the errors \cite{2018PhRvD..98b3527A}.  Furthermore, when using early- or late-time data the sound horizon is inferred as a derived parameter, specifically for early times one adopts data from the CMB anisotropic power spectra while for late times data from the galaxy correlation function are used \cite{2018PhRvD..98b3527A}. In both scenarios the data are fitted assuming an underlying cosmological model. This methodology of inferring distances from BAO is common to every survey from the Sloan Digital Sky Survey (SDSS) \cite{2000AJ....120.1579Y} to the recent DESI Collaboration \cite{2025JCAP...07..028A, 2025PhRvD.112h3515A} and it is known as \emph{BAO-Only}, adopted to overcome late times non-linearities. Specifically, the correlation function is fit with phenomenological templates with fixed value of the cosmological parameters for a flat $\Lambda$CDM model and the nonlinear damping parameters leading to possible underestimation of the errors and unclear validity to models that are not extensions of the concordance paradigm \cite{2023PhRvD.107l3506A, 2019PhRvD..99l3515A}. 

In light of this, in recent years new methodologies, under the name of Purely-Geometric-BAO (PGB) approaches to infer the distances have been proposed that do not rely on a fixed cosmology trying to make BAO a true geometric probe \cite{2019PhRvD..99l3515A, 2016MNRAS.455.2474A, 2018PhRvL.121b1302A, 2018PhRvD..98b3527A, 2020PhRvD.101h3517O, 2021JCAP...01..009P, 2023PhRvD.107l3506A}. These PGB methods have the advantage of not assuming a specific dark energy fluid and spatial curvature of the Universe in two different ways, i.e. the correlation function model fitting (CF-MF) and the linear point (LP) where the latter is used to overcome possible limits of the CF-MF \cite{2023PhRvD.107l3506A, 2019PhRvD..99l3515A}.

Focusing on the LP, considering the clustering correlation function its position is found between the peak and the dip in the correlation function and it has the advantage of being redshift independent and insensitive to the dark energy density and spatial curvature of the Universe \cite{2023PhRvD.107l3506A, 2019PhRvD..99l3515A, 2016MNRAS.455.2474A}. The strategy is to fit a differentiable function, which is model-independent, to the correlation function monopole data to directly estimate the location of the peak and dip and then derive the LP from them \cite{2020PhRvD.101h3517O}. With this method, the distances can be estimated without using a phenomenological model for the correlation function. This methodology was validated through, e.g. mock survey catalogs \cite{2018PhRvD..98b3527A} and massive neutrinos \cite{2021JCAP...01..009P}.

Interestingly, recently another paper has quantified the systematics on current BAO surveys finding linearly increasing bias with the distances when a fiducial cosmological model is adopted  \cite{2026arXiv260303443A}.

However, given the impact of such probe with investigating late-time domains we still use BAO data from the most recent survey, i.e. the DESI Collaboration \cite{2025JCAP...07..028A, 2025PhRvD.112h3515A} bearing in mind that they exhibit the criticalities highlighted here. In light of this, even if we still adopt DESI-BAO we perform also analyses where the sample is not used to assess how they impact our computations.

\section{Preliminary numerical outcomes}\label{numout}

To derive the constraints on the cosmographic parameters we first consider two preliminary analyses, labeled \emph{Analysis 1} and \emph{Analysis 2} where we find the preliminary parameters by maximizing the following log-likelihoods
\begin{align*}
     &\text{{\bf Analysis 1:}}\ \ln\mathcal{L}_S+\ln\mathcal{L}_A,\\
    &\text{{\bf Analysis 2:}}\ \ln\mathcal{L}_S+\ln\mathcal{L}_A+\ln\mathcal{L}_D.
\end{align*}
All analyses are done by modifying \texttt{CLASS} \cite{2011arXiv1104.2932L, 2011JCAP...07..034B, 2011arXiv1104.2934L} and performing a MCMC analysis using the \texttt{MontePython} sampler \cite{2019PDU....24..260B, 2013JCAP...02..001A}. The constrained parameters with attached $1$-$\sigma$($2$-$\sigma$) errors for both analyses are depicted in Tab. \ref{tab:bfcosmo}\footnote{When adopting the GRB data we decide  not to include the parameters of the correlation since they are not functional for our purposes. However, for clarity we placed them in Tab. \ref{tab:bfGRB} of Appendix \ref{calibration}.} while the contour plots generated using the Python package \texttt{GetDist} \cite{2025JCAP...08..025L} are in Fig. \ref{fig:contpre} of Appendix \ref{precont}. We adopt the Gelman-Rubin convergence criterion \cite{1992StaSc...7..457G}, requiring $R-1< 0.02$.

Furthermore, the results inferred from our computations are then compared with the expected values of $q_0$ and $j_0$ for the two cosmological models most widely discussed in the literature, namely the $\Lambda$CDM and the $\omega_0\omega_1$CDM scenarios \cite{2020A&A...641A...6P, 2001IJMPD..10..213C, 2003PhRvL..90i1301L}. 

We begin by writing the Hubble rate $H(z)$ as
\begin{equation}\label{hrate}
    H(z)=H_0\sqrt{\Omega_m(1+z)^3+\Omega_{de}(z)},
\end{equation}
where $H_0$ is the Hubble constant whereas $\Omega_m$ and $\Omega_{de}$ are the matter and dark energy parameters, respectively. The  case of a pure cosmological constant requires that $\Omega_{de}(z)\equiv 1-\Omega_m$ for all the redshifts.

To write the cosmographic parameters in terms of the quantities entering the concordance or $\omega_0\omega_1$CDM models, we consider the relation between $q(z)$, $j(z)$ and $H(z)$ 
\begin{subequations}\label{qjrel}
    \begin{align}
        &q(z)=-1+(1+z)\frac{H^\prime}{H},\\
        &j(z)=1-(1+z)\frac{H^\prime}{H}\left[2-(1+z)\left(H^{\prime 2}+\frac{H^{\prime\prime}}{H^\prime}\right)\right],
    \end{align}
\end{subequations}
where the $\prime$ denotes the derivative with respect to the redshift. Substituting Eq. \eqref{hrate} inside Eqs. \eqref{qjrel} and considering that at the present epoch $z=0$ we end up with the expression for the deceleration and jerk parameters for the cosmological models of our interest.

\begin{itemize}
    \item [-] {\bf $\Lambda$CDM model.} Starting from the concordance paradigm, following the prescription illustrated above the deceleration and jerk parameters at the present epoch for the $\Omega_k=0$ case are 
    \begin{subequations}
        \begin{align}
            &q_0=\frac{3}{2}\Omega_m-1,\\
            &j_0=1.
        \end{align}
    \end{subequations}
    Now, considering that $\Omega_m=0.3153\pm0.0073$ \cite{2020A&A...641A...6P} we end up with $q_0=-0.527\pm0.011$.
    \item [-] {\bf $\omega_0\omega_1$CDM model.} Now, for the case of dynamical dark energy we end up with the following expressions for $q_0$ and $j_0$
    \begin{subequations}
        \begin{align}
            &q_0=\frac{1}{2}(1+3\omega_0\left(1-\Omega_m\right)),\\
            &j_0=1+\frac{1}{2}(3(1-\Omega_m)(\omega_1+3\omega_0(1+\omega_0))).
        \end{align}
    \end{subequations}
    Also in this case we derive the values of the deceleration and jerk parameters adopting $\omega_0=-0.957\pm 0.080$ and $\omega_1=-0.29^{+0.32}_{-0.26}$ \cite{2020A&A...641A...6P} ending up with $q_0=-0.483\pm 0.093$ and $j_0=0.575^{+0.558}_{-0.497}$.
\end{itemize}

\begin{table}[t]
\scriptsize
\centering
\setlength{\tabcolsep}{1.em}
\renewcommand{\arraystretch}{1.3}
\begin{tabular}{ccc}
\hline\hline
 $H_0$ & $q_0$ & $j_0$ \\
\hline
\multicolumn{3}{c}{{\bf Taylor}}\\
\hline
\multicolumn{3}{c}{{\it Analysis 1}}\\
\cline{1-3}
$73.47^{+0.889(2.063)}_{-1.076(1.950)}$  & $-0.436^{+0.058(0.113)}_{-0.056(0.113)}$ & $0.605^{+0.211(0.469)}_{-0.257(0.447)}$\\
\hline
\multicolumn{3}{c}{{\it Analysis 2}}\\
\cline{1-3}
$69.20^{+0.568(1.068)}_{-0.547(1.100)}$ & $-0.485^{+0.028(0.055)}_{-0.028(0.056)}$ & $0.683^{+0.048(0.098)}_{-0.051(0.097)}$
\\
\hline
\multicolumn{3}{c}{\bf Pad\'e}\\
\hline
\multicolumn{3}{c}{\it Analysis 1}\\
\cline{1-3}
$73.59^{+1.006(1.985)}_{-0.985(2.025)}$ & $-0.501^{+0.088(0.165)}_{-0.073(0.177)}$ & $1.269^{+0.470(1.397)}_{-0.723(1.241)}$\\
\hline
\multicolumn{3}{c}{{\it Analysis 2}}\\
\cline{1-3}
$69.55^{+0.567(1.105)}_{-0.546(1.122)}$ & $-0.629^{+0.051(0.098)}_{-0.048(0.100)}$ & $1.931^{+0.255(0.580)}_{-0.310(0.552)}$\\
\hline\hline
\end{tabular}
\caption{Mean values of the cosmographic parameters with attached $1$-$\sigma$($2$-$\sigma$) errors in \emph{Analysis 1}-\emph{Analysis 2}. Upper panel shows the results when the Taylor expansion is adopted while the lower panel depicts the results for the Pad\'e approximation. $H_0$ is in units of $\mathrm{km\,s^{-1}\,Mpc^{-1}}$.}
\label{tab:bfcosmo}
\end{table}

\subsection{Analysis 1 results}

We discuss the preliminary outcomes of our MCMC analyses starting from \emph{Analysis 1} where we employ the Pantheon+\&SH0ES samples \cite{2022ApJ...938..113S, 2022ApJ...938..110B} together with GRB data following the $E_p-E_{iso}$ correlation function \cite{2021JCAP...09..042K}.

\begin{itemize}
    \item [-] {\bf Taylor.} Our preliminary results suggest agreement at the $1\sigma$ level between our value of $H_0$ and $H_0^R=(73.04\pm1.04),\mathrm{km,s^{-1},Mpc^{-1}}$ \cite{2022ApJ...934L...7R} while no agreement has been found with $H^P_0=(67.36\pm 0.54)\ \text{km/s/Mpc}$ \cite{2020A&A...641A...6P}. On the other hand, the deceleration parameter $q_0$ agrees at $1$-$\sigma$ with the $\omega_0\omega_1$CDM and \emph{only} at $2$-$\sigma$ with the concordance paradigm. This behavior is also exhibited in the jerk parameter $j_0$ which, analogously to $q_0$, agrees at $1$-$\sigma$ with a dynamical dark energy scenario and only at $2$-$\sigma$ with the cosmological constant.
    
    \item [-] {\bf Pad\'e (2,1).} Alleviating the convergence issues of the Taylor expansion also changes our results for $q_0$. Specifically, the compatibility with $\Lambda$CDM rises at $1$-$\sigma$ like for the $\omega_0\omega_1$CDM scenario. As for the Hubble constant $H_0$ and the jerk parameter $j_0$ they still remain compatible at $1$-$\sigma$ only with $H^R_0$ \cite{2022ApJ...934L...7R} and at $1$-$\sigma$($2$-$\sigma$) with the $\omega_0\omega_1$CDM($\Lambda$CDM) models, respectively.
    
\end{itemize}

\subsection{Analysis 2 results}

We discuss the preliminary outcomes of our MCMC analyses starting from \emph{Analysis 2} where we employ the same datasets of \emph{Analysis 1} but with the addition of the DESI-BAO \cite{2025PhRvD.112h3515A}.

\begin{itemize}
    \item [-] {\bf Taylor.} The addition of the BAO sample leads to a degradation of our results of the Hubble constant $H_0$ and the jerk parameter $j_0$ since, concerning $H_0$, now it is not in agreement with either $H^P_0$ or $H^R_0$ while $j_0$ does not agree with the prediction of both the $\Lambda$CDM and $\omega_0\omega_1$CDM models. On the other hand, for the deceleration parameter $q_0$ we found the same as for \emph{Analysis 1}, i.e. it agrees at $1$-$\sigma$ with the $\omega_0\omega_1$CDM and at $2$-$\sigma$ with the concordance paradigm.
    \item [-] {\bf Padé (2,1).} When Pad\'e is adopted, the worsening of our results is more pronounced than in the Taylor case. Also here, $H_0$ is in agreement with neither $H_0^P$ from the Planck Collaboration \cite{2020A&A...641A...6P} nor $H^R_0$ from  \citet{2022ApJ...934L...7R}. In contrast with dropping the DESI sample, $j_0$ shows no agreement with those inferred from the cosmological scenarios discussed while the agreement with $q_0$ drops at $2$-$\sigma$ for both the concordance and $\omega_0\omega_1$CDM scenarios.
\end{itemize}

As already discussed earlier, other works have found tensions between the jerk parameter inferred from DESI together with various dataset combinations. \citet{2024A&A...690A..40L} found disagreement between the $j_0$ inferred when using DESI-DR1+OHD+SNeIa and the one inferred when considering $\Lambda$CDM or $\omega_0\omega_1$CDM models. This tension was further investigated by \citet{2025PhRvD.112j3519R} finding maximum deviation from $1$ at $5.4\sigma$ when DESI-DR2 is combined with the DESY5-SNeIa sample.

\begin{figure*}[t]
    \centering
    \includegraphics[width=.5\linewidth]{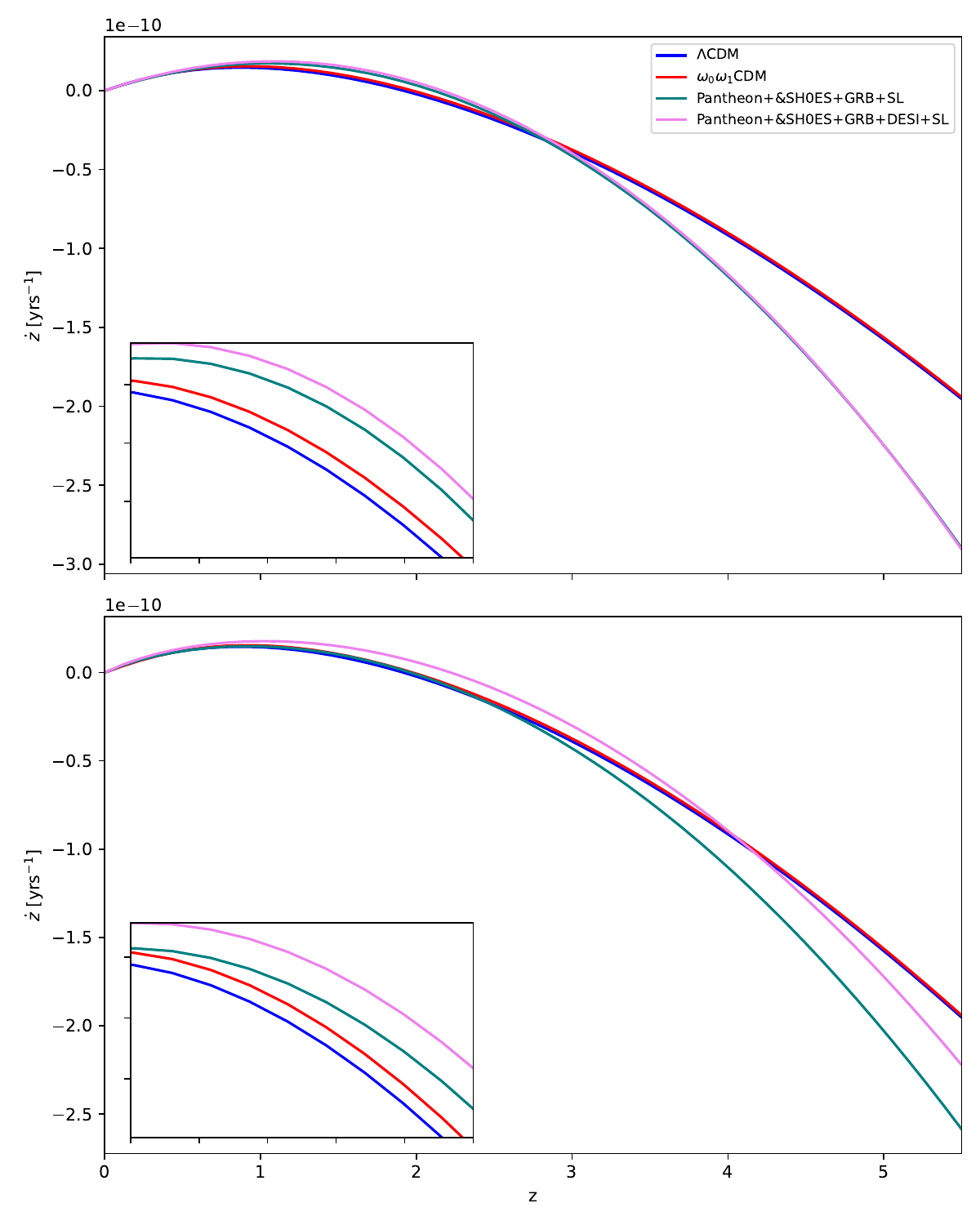}
    \caption{Behavior of the redshift drift $\Dot{z}$ in units of $\text{yrs}^{-1}$ for the flat $\Lambda$CDM and $\omega_0\omega_1$CDM models compared with $\Dot{z}$ written adopting the cosmographic approach in both cases of Taylor (upper plot) and Pad\'e (lower plot). The values of $H_0$, $q_0$ and $j_0$ are taken from Tab. \ref{tab:bfsl} for \emph{Analysis 1SL-2SL}. For the $\Lambda$CDM and $\omega_0\omega_1$CDM models we consider $H_0$, $\Omega_m$, $\omega_0$ and $\omega_1$ from the Planck Collaboration \cite{2020A&A...641A...6P}. In both figures on the bottom left a zoom plot in the interval $z\in [1,\ 2]$.}
    \label{fig:zdriftflat}
\end{figure*}

\section{Constraints with the Sandage-Loeb test}\label{mocksl}

Here we derive the new constraints on the cosmographic parameters when the SL datasets are taken into account. In this case we label \emph{Analysis 1SL} when the mock SL data are employed. Also in this case the Gelman-Rubin convergence criterion \cite{1992StaSc...7..457G} is adopted, requiring $R-1\lesssim 0.04$. The constrained parameters in this case are derived by maximizing the following log-likelihoods
\begin{align*}
     &\text{{\bf Analysis 1SL:}}\ \ln\mathcal{L}_S+\ln\mathcal{L}_A+\ln\mathcal{L}_{SL},\\
    &\text{{\bf Analysis 2SL:}}\ \ln\mathcal{L}_S+\ln\mathcal{L}_A+\ln\mathcal{L}_D+\ln\mathcal{L}_{SL}.
\end{align*}
As in Sect. \ref{numout}, also in this case the computations are performed adopting both the \texttt{CLASS} \cite{2011arXiv1104.2932L, 2011JCAP...07..034B, 2011arXiv1104.2934L} and \texttt{MontePython} \cite{2019PDU....24..260B, 2013JCAP...02..001A} codes with the confidence contours created through \texttt{GetDist} \cite{2025JCAP...08..025L}.

\begin{figure*}[t]
    \centering

    \begin{minipage}{0.44\textwidth}
        \centering
        \includegraphics[width=\linewidth]{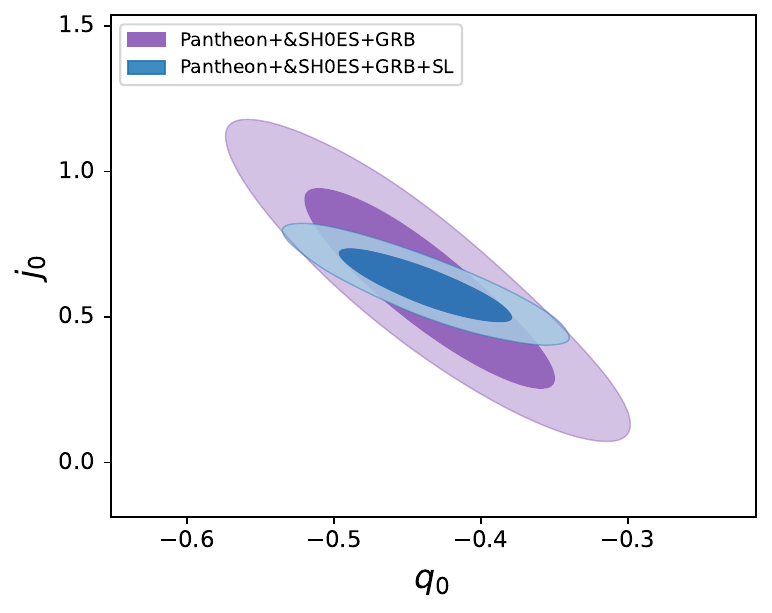}
    \end{minipage}
    \hfill
    \begin{minipage}{0.44\textwidth}
        \centering
        \includegraphics[width=\linewidth]{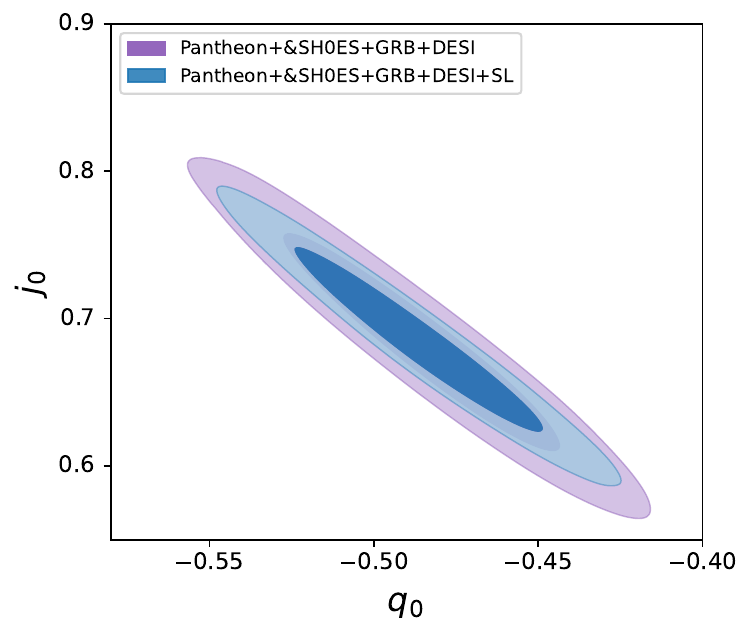}
    \end{minipage}

    \vspace{0.3cm}

    \begin{minipage}{0.44\textwidth}
        \centering
        \includegraphics[width=\linewidth]{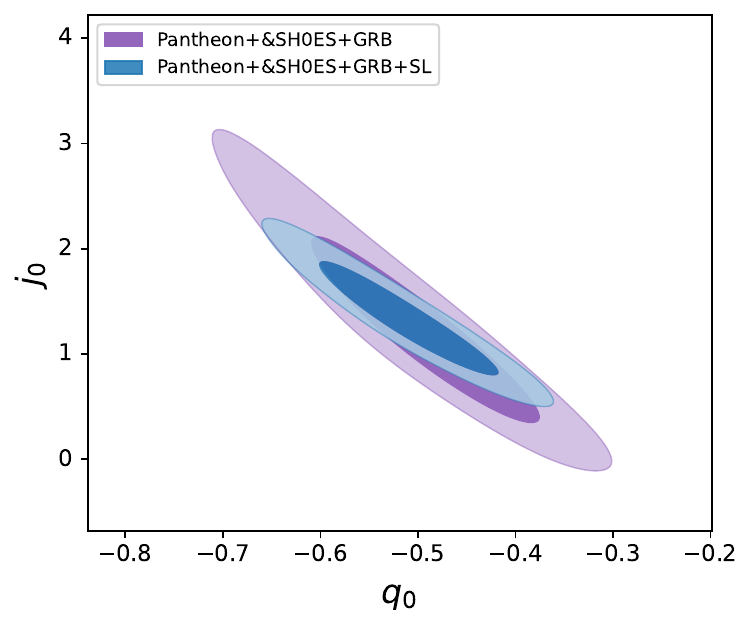}
    \end{minipage}
    \hfill
    \begin{minipage}{0.44\textwidth}
        \centering
        \includegraphics[width=\linewidth]{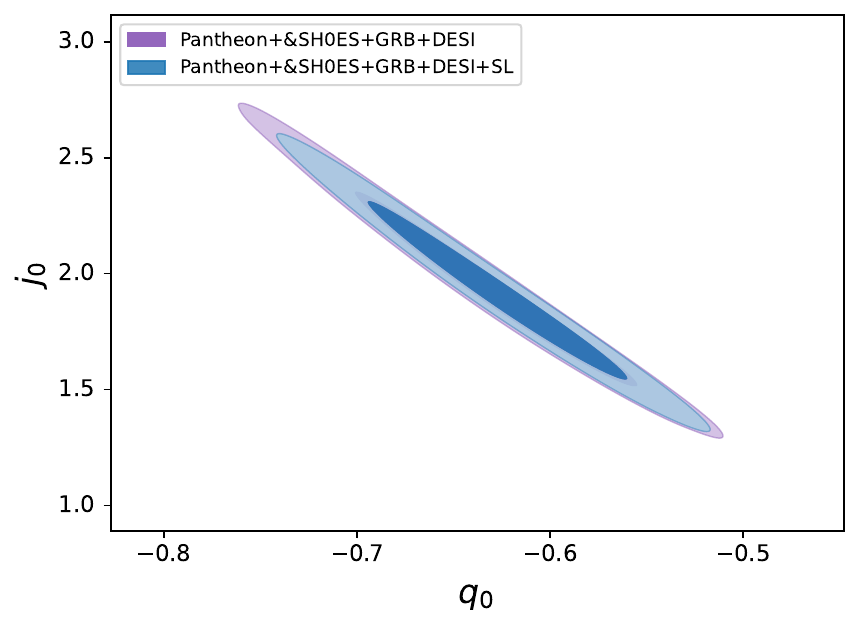}
    \end{minipage}

    \caption{Confidence contours representation comparing \emph{Analyses 1-2} with \emph{Analyses 1SL-2SL} for the $q_0-j_0$ plane. Upper panel shows the contours when the Taylor series is adopted with (upper right) and without (upper left) DESI-BAO while lower panel shows them when the $(2,1)$ Pad\'e approximation is used with (lower right) and without (lower left) DESI-BAO.}
    
    \label{fig:contSL}
\end{figure*}

The updated parameters with attached $1$-$\sigma$($2$-$\sigma$) error bars are illustrated in Tab. \ref{tab:bfsl} and their confidence contours, compared with the ones from \emph{Analyses 1-2}, in Fig. \ref{fig:contSL}. As can be seen from the contours and from Tabs. \ref{tab:bfcosmo}-\ref{tab:bfsl}, introducing the SL mock data in our computations leads to a reduction in the uncertainties on the deceleration and jerk parameters.

\begin{table}[H]
\scriptsize
\centering
\setlength{\tabcolsep}{1.em}
\renewcommand{\arraystretch}{1.3}
\begin{tabular}{cccc}
\hline\hline
 $H_0$ & $q_0$ & $j_0$ \\
\hline
\multicolumn{3}{c}{{\bf Taylor}}\\
\hline
\multicolumn{3}{c}{{\it Analysis 1SL}}\\
\cline{1-3}
$73.39^{+0.977(2.026)}_{-1.030(2.001)}$  & $-0.438^{+0.040(0.082)}_{-0.040(0.081)}$ & $0.607^{+0.085(0.177)}_{-0.089(0.174)}$\\
\hline
\multicolumn{3}{c}{{\it Analysis 2SL}}\\
\cline{1-3}
$69.24^{+0.514(1.042)}_{-0.532(1.038)}$ & $-0.486^{+0.026(0.052)}_{-0.026(0.053)}$ & $0.685^{+0.041(0.088)}_{-0.045(0.085)}$
\\
\hline
\multicolumn{3}{c}{\bf Pad\'e}\\
\hline
\multicolumn{3}{c}{\it Analysis 1SL}\\
\cline{1-3}
$73.65^{+1.022(2.007)}_{-0.998(2.027)}$ & $-0.512^{+0.062(0.124)}_{-0.062(0.125)}$ & $1.339^{+0.340(0.748)}_{-0.393(0.714)}$
\\
\hline
\multicolumn{3}{c}{\it Analysis 2SL}\\
\cline{1-3}
$69.51^{+0.500(1.074)}_{-0.520(1.039)}$ & $-0.628^{+0.048(0.087)}_{-0.040(0.094)}$ & $1.923^{+0.215(0.550)}_{-0.284(0.489)}$
\\
\hline\hline
\end{tabular}
\caption{Mean values of the cosmographic parameters with attached $1$-$\sigma$($2$-$\sigma$) errors in \emph{Analyses 1SL-2SL}. Upper panel shows the results when the Taylor expansion is adopted while the lower panel depicts the results for the Pad\'e approximation. $H_0$ is in units of $\mathrm{km\,s^{-1}\,Mpc^{-1}}$.}
\label{tab:bfsl}
\end{table}

The constraints on the cosmographic parameters with the addition of the SL probe do not change much from \emph{Analyses 1-2} with some exceptions.

\begin{itemize}
    \item [-] When considering \emph{Analysis 1} with the addition of the SL, i.e. \emph{Analysis 1SL} when using Taylor we find that the compatibility between our $q_0$ and $q_0=-0.483\pm 0.093$ considering the $\omega_0\omega_1$CDM model decreases at $2$-$\sigma$. Also, our $j_0$ does not agree anymore with $j_0=1$ from the concordance paradigm. On the other hand, using Pad\'e improves the agreement between $j_0$ and $1$ at $1$-$\sigma$.
    
    \item [-] Considering the DESI sample together with the SL, i.e. \emph{Analysis 2SL} in the case of $H(z)$ written in terms of Pad\'e polynomials the deceleration parameter is only compatible at $2$-$\sigma$ with the $\omega_0\omega_1$CDM scenario while for \emph{Analysis 2} the compatibility at $2$-$\sigma$ was also found for the concordance paradigm. 
\end{itemize}

This change in compatibility in \emph{Analysis 2} when adding the mock SL is reflected mainly in the deceleration parameter $q_0$ when DESI is used. This should not be a surprise, since the SL signal is a direct probe of the acceleration or deceleration of the Universe with the deceleration parameter $q_0$ directly encoding this behavior. However, it is worth noticing that when the BAO are dropped, i.e. \emph{Analysis 1SL} also the jerk parameter $j_0$ is affected.

Moreover, as can be clearly seen in Fig. \ref{fig:contSL} adding SL mock data gives tighter constraints on the $q_0-j_0$ plane especially when BAO data are not used, i.e. upper and lower left. The shrinkage observed in the contours should not be a surprise since our mock catalog is generated considering both the same $H(z)$ from the preliminary fits and the cosmographic parameters inferred from \emph{Analysis 1-2}.

Also, focusing on the behavior of $\Dot{z}$ depicted in Fig. \ref{fig:zdriftflat} we find that when Taylor is used (upper plot) the behavior of the drift when using SNeIa+GRB+SL and SNeIa+GRB+DESI+SL follows $\Dot{z}$ written in terms of the concordance or $\omega_0\omega_1$CDM until $z\approx 3.2$. On the other hand, when $H(z)$ is written using Pad\'e we observe that until $z\approx 3$ the curve inferred from SNeIa+GRB+SL follows the ones for both cosmological models while the curve from SNeIa+GRB+DESI+SL seems to be more detached from them until $z\approx 4$. Finally, zooming in on both plots, we can clearly see that, especially with Pad\'e, the curve of $\dot z$ for the SNeIa+GRB+SL case seems to follow more closely that of the $\omega_0\omega_1$CDM model than that of the $\Lambda$CDM scenario.

However, this may also be due to the use of GRB data. For example, \citet{2020A&A...641A.174L} showed that adopting different $H(z)$ written in terms of Taylor, Pad\'e and auxiliary variables with calibrated GRB correlation functions alone, the constraints on $q_0$ and $j_0$ are not compatible with the $\Lambda$CDM model. Thus, this trend may be driven principally by the presence of GRB in both analyses.

\section{Conclusions}\label{conc}

In this work, we investigated the redshift drift within a cosmographic framework by combining current observational probes with a mock SL dataset. The analysis was performed by modeling the Hubble rate through both a second order Taylor expansion and a Pad\'e $(2,1)$ approximation and by constraining the cosmographic parameters, $H_0$, $q_0$, and $j_0$, by virtue of a MCMC approach.

We first derived constraints using two complementary datasets. The first analysis, based on SNeIa and GRB data, provided results compatible at the $1\sigma$ level with a higher value of the Hubble constant, while no agreement was found with lower values inferred from CMB observations. In this case, the inferred deceleration and jerk parameters favored a dynamical dark energy scenario when the Taylor expansion was adopted, while the Pad\'e parameterization increased the compatibility with the concordance paradigm, particularly for the deceleration parameter.

The inclusion of DESI BAO data in the second analysis led to a degradation of the overall consistency with reference cosmological models. In particular, the inferred Hubble constant showed no agreement with either local or CMB determinations and the jerk parameter did not match the predictions of both $\Lambda$CDM and dynamical dark energy scenarios. The deceleration parameter remained consistent with previous results when Taylor was used, while its compatibility weakened when adopting Pad\'e. The role of DESI BAO is also critically revised and naively discussed throughout the manuscript.

Afterwards, we constructed a mock SL dataset using the cosmographic parameters inferred from the preliminary analyses and repeated the MCMC computations including this additional probe. This procedure did not constitute an independent forecast, but rather a robust consistency test within the same cosmographic framework.

Accordingly, the inclusion of the SL mock data led to a significant reduction of the uncertainties on the deceleration and jerk parameters with only moderate shifts in their central values. The impact of the SL signal was found to be primarily on the deceleration parameter when BAO data were included, reflecting the direct sensitivity of the redshift drift to the acceleration of the Universe. When BAO data were excluded, variations were also observed in the jerk parameter, indicating a stronger interplay between high-redshift probes and higher-order cosmographic terms.

The comparison between the reconstructed redshift drift and the predictions from reference cosmological models showed that the cosmographic reconstruction generally followed the expected behavior up to intermediate redshifts. However, deviations appeared at higher redshift, particularly when DESI data were included and when Pad\'e parameterizations were adopted. In contrast, the reconstruction based on SNeIa and GRB data alone showed a closer agreement with a dynamical dark energy scenario.

Overall, our results indicated that the SL signal is effective in tightening constraints on cosmographic parameters, especially those related to the acceleration history. However, its impact on the central values remained limited within the adopted framework. The analysis also highlighted a direct dependence on the choice of datasets and parameterization of the Hubble rate, as well as a sensitivity to the inclusion of BAO and GRB data.

For the sake of completeness, we remarked that the construction of the mock SL catalog, from internally inferred cosmographic parameters, naturally led to the observed improvement in the constraints, reinterpreted as an internal consistency check of the entire set of analyses, here presented.  

Future investigations will aim at assessing the robustness of these results by adopting alternative parameterizations of the Hubble rate and by constructing mock datasets from external fiducial cosmologies. In addition, the role of GRB data in driving deviations from the concordance paradigm will be further explored, as well as the impact of independent BAO reconstructions and additional SL measurements probing different redshift regimes.

\section*{Acknowledgements}

ACA acknowledges the support of the Istituto Nazionale di Fisica Nucleare (INFN) Sezione di Napoli, Iniziativa Specifica QGSKY. 
This paper is based upon work from COST Action CA21136 - Addressing observational tensions in cosmology with systematics and fundamental physics (CosmoVerse), supported by COST (European Cooperation in Science and Technology). The authors are also very grateful to Olga Mena for kind support and  hospitality of ACA during the time in which this manuscript has been finalized. OL is very thankful to Stefano Anselmi and Alessandro Renzi for interesting discussions on BAOs and their applications in modern cosmology. 

\bibliography{bibliography}

\newpage

\begin{appendices}

\section{Calibration of the Amati correlation}\label{calibration}

This appendix is dedicated to the illustration of the method adopt to standardize the GRB sample. As already mentioned, the $E_p-E_{iso}$ correlation is model-dependent through the luminosity distance $d_L(z)$ inside the isotropic radiated energy $E_{iso}$. Thus, our recipe consists in considering a reconstructed Hubble rate via B\'ezier polynomials up to $n$-order applied to the OHD data catalog
\begin{equation}\label{Hzbezier}
    H_n(y)= g_\alpha\sum^{n}_{i=0}\alpha_i\frac{n!}{i!}\frac{y^i(1-y)^{n-i}}{(n-i)!}\ \text{km/s/Mpc},
\end{equation}

where $g_\alpha\equiv 100$, $0\leq y\equiv z/z_M \leq 1$ with $z_M=1.965$, the maximum redshift of the OHD dataset and $\alpha_i=\{\alpha_0, \alpha_1, \alpha_2\}$ are the coefficients of the B\'ezier curve with $\alpha_0\equiv h_0$. For our purposes we stop at $n=2$ since it is the only order exhibiting a non-linear monotonic behavior \cite{2019MNRAS.486L..46A, 2024A&A...686A..30A}.

The first step in our calibration is to derive the $\alpha_i$ B\'ezier coefficients. To do that we modify the \texttt{CLASS} code \cite{2011arXiv1104.2932L, 2011JCAP...07..034B, 2011arXiv1104.2934L} and perform a MCMC analysis using the \texttt{MontePython} sampler \cite{2019PDU....24..260B, 2013JCAP...02..001A} by maximizing the following log-likelihood
\begin{equation}
    \ln\mathcal{L}_O\propto-\frac{1}{2}\sum^{N_O}_{i=1}\left[\frac{H_i-H_2(z_i)}{\sigma^2_{H_i}}\right]^2,
\end{equation}
where $H_i$ are the OHD data points with attached $\sigma_{H_i}$ errors and $N_O=34$ the adopted data points displayed in Tab. I of Ref. \cite{2026JHEAp..4900444A}.

\begin{figure}[h!]
    \centering
    \includegraphics[width=0.85\linewidth]{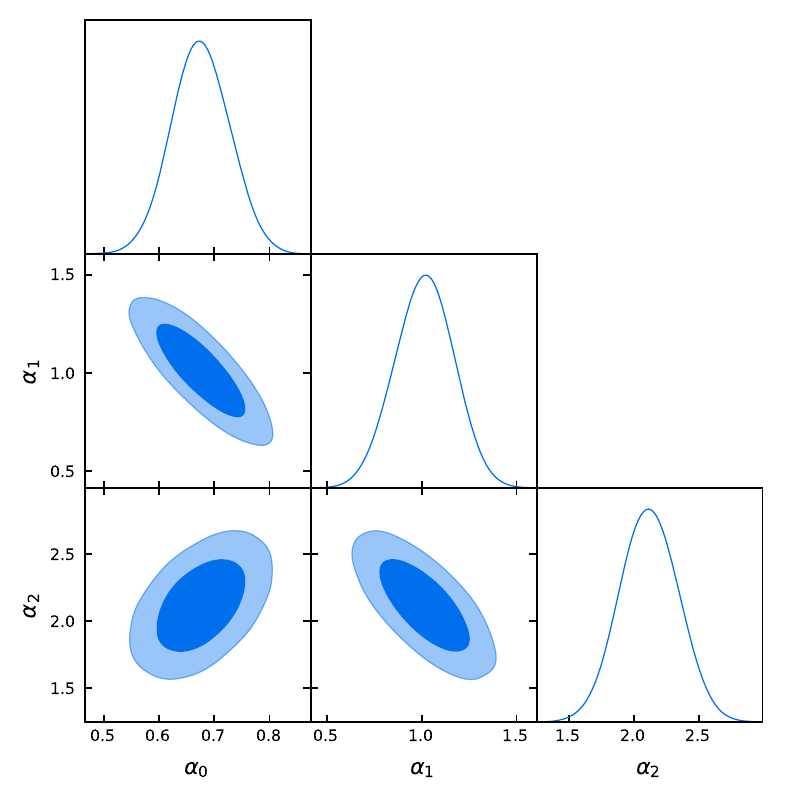}
    \caption{Contour plot of the B\'ezier coefficients. Darker (lighter) blue areas depict the $1$-$\sigma$($2$-$\sigma$) confidence level.}
    \label{fig:Bezier}
\end{figure}

The contour plot portrayed in Fig. \ref{fig:Bezier} is generated using the \texttt{GetDist} package \cite{2025JCAP...08..025L} while the constrained parameters with attached $1$-$\sigma$($2$-$\sigma$) errors inferred from our computations are
\begin{subequations}\label{bezier}
    \begin{align}
    \alpha_0=0.675^{+0.055(0.111)}_{-0.057(0.111)},\\
    \alpha_1=1.015^{+0.166(0.322)}_{-0.160(0.322)},\\
    \alpha_2=2.117^{+0.234(0.472)}_{-0.239(0.461)}.
    \end{align}
\end{subequations}
Now, we are able to adopt the $E_p-E_{iso}$ correlation in a total model-independent way. The next step is to substitute the inferred B\'ezier coefficients in Eq. \eqref{bezier} inside the luminosity distance in the isotropic radiated energy $E_{iso}$ through Eq. \eqref{Hzbezier} at $n=2$ leading to
\begin{equation}
    d_L(z)=c(1+z)\int^z_0 \frac{dz^\prime}{H_2(z^\prime)}.
\end{equation}
This will give us the calibrated GRB sample enabling us to use them as distance indicators.

We also illustrate the GRB parameters inferred from both \emph{Analysis 2} and \emph{Analysis 2SL}. In the following Table the mean values of the parameters with attached $1$-$\sigma$($2$-$\sigma$) errors are presented

\begin{table}[H]
\scriptsize
\centering
\setlength{\tabcolsep}{0.18em}
\renewcommand{\arraystretch}{1.6}
\begin{tabular}{@{} l c c c @{}}
\hline\hline
 & $a$ & $b$ & $\sigma$ \\
\hline
\multicolumn{4}{c}{\it Analysis 1} \\
\hline
Taylor & $0.743^{+0.037(0.083)}_{-0.043(0.079)}$ & $1.804^{+0.061(0.114)}_{-0.053(0.119)}$ & $0.301^{+0.021(0.047)}_{-0.025(0.044)}$ \\
Pad\'e & $0.744^{+0.037(0.080)}_{-0.042(0.077)}$ & $1.804^{+0.059(0.108)}_{-0.052(0.114)}$ & $0.301^{+0.020(0.045)}_{-0.024(0.043)}$  \\
\hline
\multicolumn{4}{c}{\it Analysis 2} \\
\hline
Taylor & $0.742^{+0.037(0.080)}_{-0.042(0.080)}$ & $1.774^{+0.037(0.114)}_{-0.054(0.116)}$ & $0.300^{+0.020(0.047)}_{-0.025(0.043)}$ \\
Pad\'e & $0.742^{+0.037(0.082)}_{-0.043(0.079)}$ & $1.774^{+0.060(0.115)}_{-0.054(0.115)}$ & $0.300^{+0.020(0.047)}_{-0.025(0.044)}$  \\
\hline
\multicolumn{4}{c}{\it Analysis 1SL} \\
\hline
Taylor & $0.745^{+0.038(0.082)}_{-0.043(0.080)}$ & $1.802^{+0.059(0.114)}_{-0.054(0.116)}$ & $0.302^{+0.020(0.048)}_{-0.025(0.044)}$  \\
Pad\'e & $0.745^{+0.037(0.082)}_{-0.044(0.080)}$ & $1.806^{+0.058(0.112)}_{-0.054(0.113)}$ & $0.302^{+0.020(0.046)}_{-0.025(0.044)}$ \\
\hline
\multicolumn{4}{c}{\it Analysis 2SL} \\
\hline
Taylor & $0.742^{+0.037(0.081)}_{-0.042(0.077)}$ & $1.773^{+0.059(0.112)}_{-0.053(0.115)}$ & $0.300^{+0.020(0.046)}_{-0.026(0.043)}$  \\
Pad\'e & $0.740^{+0.037(0.081)}_{-0.042(0.079)}$ & $1.778^{+0.060(0.112)}_{-0.054(0.116)}$ & $0.299^{+0.019(0.046)}_{-0.024(0.042)}$ \\
\hline\hline
\end{tabular}
\caption{Mean values of the GRB correlation parameters with attached $1$-$\sigma$($2$-$\sigma$) error bars. Upper panels show the results for \emph{Analysis 1}-\emph{Analysis 2} when using Taylor or Pad\'e approximation of $H(z)$. Lower panels show the same but for \emph{Analysis 1SL}-\emph{Analysis 2SL}.}
\label{tab:bfGRB}
\end{table}

\onecolumngrid

\section{Preliminary contour plots}\label{precont}

We display the contour plots for the preliminary numerical outcomes. Specifically, we show the contours on the GRB correlation and cosmographic parameters, namely the Hubble constant $H_0$, deceleration $q_0$ and jerk $j_0$ parameters for the Taylor or Pad\'e parameterizations of the Hubble rate $H(z)$ for \emph{Analysis 1} and \emph{Analysis 2}.

\begin{figure}[h!]
    \centering
    \begin{minipage}{0.47\textwidth}
        \centering
        \includegraphics[width=\linewidth]{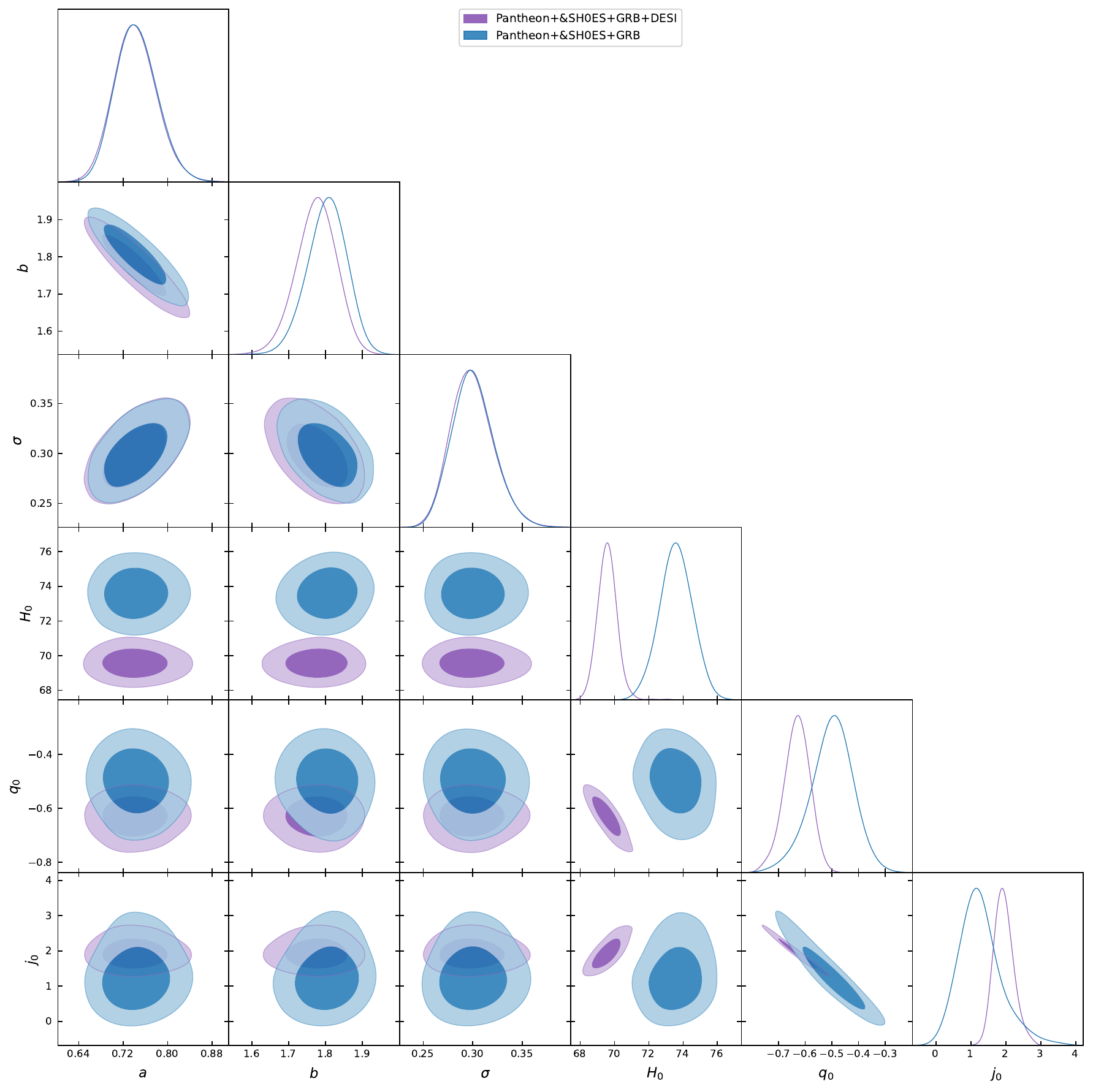}
    \end{minipage}
    \begin{minipage}{0.47\textwidth}
        \centering
        \includegraphics[width=\linewidth]{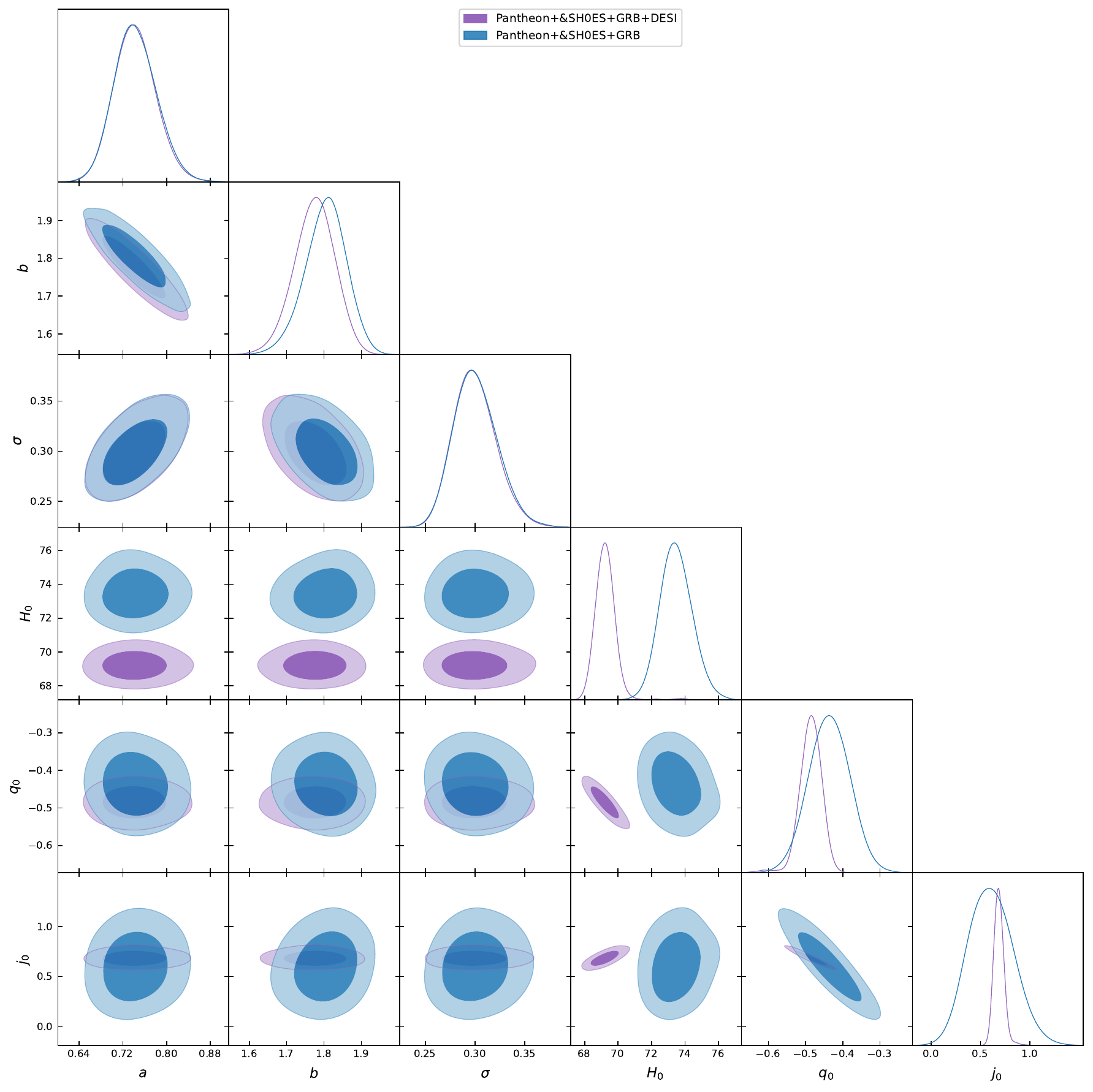}
    \end{minipage}
    \caption{Contour plots for the preliminary MCMC outcomes for the GRB and cosmographic parameters in both the case of the Taylor or Pad\'e series for \emph{Analysis 1}-\emph{Analysis 2}. Right panel shows the contour plot when Taylor is used while the left panel shows the contour plot when using Pad\'e.}
    \label{fig:contpre}
\end{figure}

\end{appendices}

\end{document}